\documentstyle[twoside,epsf]{article}
 
\catcode`\@=11
\long\def\@makefntext#1{ 
\protect\noindent \hbox to 3.2pt {\hskip-.9pt
$^{{\eightrm\@thefnmark}}$\hfil}#1\hfill} 
 
\def\thefootnote{\fnsymbol{footnote}}
 \def\@makefnmark{\hbox to 0pt{$^{\@thefnmark}$\hss}}  
 
\def\ps@myheadings{\let\@mkboth\@gobbletwo
\def\@oddhead{\hbox{} 
\rightmark\hfil\eightrm\thepage}
\def\@oddfoot{}\def\@evenhead{\eightrm\thepage\hfil 
\leftmark\hbox{}}\def\@evenfoot{}
\def\sectionmark##1{}\def\subsectionmark##1{}}
 
\oddsidemargin=\evensidemargin
\renewcommand{\thefootnote}{\fnsymbol{footnote}}

\newcounter{sectionc}\newcounter{subsectionc}\newcounter{subsubsectionc}
\renewcommand{\section}[1] {\vspace{12pt}\addtocounter{sectionc}{1}
\setcounter{subsectionc}{0}\setcounter{subsubsectionc}{0}\noindent
        {\tenbf\thesectionc. #1}\par\vspace{5pt}}
\renewcommand{\subsection}[1] {\vspace{12pt}\addtocounter{subsectionc}{1}
        \setcounter{subsubsectionc}{0}\noindent
        {\bf\thesectionc.\thesubsectionc. {\kern1pt \bfit #1}}\par\vspace{5pt}}
\renewcommand{\subsubsection}[1] {\vspace{12pt}\addtocounter{subsubsectionc}{1}
        \noindent{\tenrm\thesectionc.\thesubsectionc.\thesubsubsectionc.
        {\kern1pt \tenit #1}}\par\vspace{5pt}}
\newcommand{\nonumsection}[1] {\vspace{12pt}\noindent{\tenbf #1}
        \par\vspace{5pt}}
 
\newcounter{appendixc}
\newcounter{subappendixc}[appendixc]
\newcounter{subsubappendixc}[subappendixc]
\renewcommand{\thesubappendixc}{\Alph{appendixc}.\arabic{subappendixc}}
\renewcommand{\thesubsubappendixc}
        {\Alph{appendixc}.\arabic{subappendixc}.\arabic{subsubappendixc}}
 
\renewcommand{\appendix}[1] {\vspace{12pt}
        \refstepcounter{appendixc}
        \setcounter{figure}{0}
        \setcounter{table}{0}
        \setcounter{lemma}{0}
        \setcounter{theorem}{0}
        \setcounter{corollary}{0}
        \setcounter{definition}{0}
        \setcounter{equation}{0}
        \renewcommand{\thefigure}{\Alph{appendixc}.\arabic{figure}}
        \renewcommand{\thetable}{\Alph{appendixc}.\arabic{table}}
        \renewcommand{\theappendixc}{\Alph{appendixc}}
        \renewcommand{\thelemma}{\Alph{appendixc}.\arabic{lemma}}
        \renewcommand{\thetheorem}{\Alph{appendixc}.\arabic{theorem}}
        \renewcommand{\thedefinition}{\Alph{appendixc}.\arabic{definition}}
        \renewcommand{\thecorollary}{\Alph{appendixc}.\arabic{corollary}}
        \renewcommand{\theequation}{\Alph{appendixc}.\arabic{equation}}
        \noindent{\tenbf Appendix \theappendixc #1}\par\vspace{5pt}}
\newcommand{\subappendix}[1] {\vspace{12pt}
        \refstepcounter{subappendixc}
        \noindent{\bf Appendix \thesubappendixc. {\kern1pt \bfit #1}}
        \par\vspace{5pt}}
\newcommand{\subsubappendix}[1] {\vspace{12pt}
        \refstepcounter{subsubappendixc}
        \noindent{\rm Appendix \thesubsubappendixc. {\kern1pt \tenit #1}}
        \par\vspace{5pt}}
 
\newcommand{\textlineskip}{\baselineskip=13pt}
\newcommand{\smalllineskip}{\baselineskip=10pt}
 
\def\eightcopyright{\copyright}
 
\newcommand{\copyrightheading}[1]
        {\vspace*{-2.5cm}\smalllineskip{\flushleft
        {\eightrm International Journal of Modern Physics C, #1}\\
        {\eightrm $\eightcopyright$\, World Scientific Publishing
         Company}\\
         }}
 
\newcommand{\publisher}[2]{{\begin{center}\eightrm\smalllineskip
        Received #1\\
        Revised #2
        \end{center}
        }}
 
\def\abstracts#1#2#3{{
        \centering{\begin{minipage}{4.5in}\baselineskip=10pt\eightrm
        \parindent=0pt #1\par
        \parindent=15pt #2\par
        \parindent=15pt #3
        \end{minipage} }\par}}
 
\renewenvironment{thebibliography}[1]                   
        {\ninerm
         \baselineskip=11pt                             
         \begin{list}{\arabic{enumi}.}
        {\usecounter{enumi}\setlength{\parsep}{0pt}
         \setlength{\leftmargin 17pt}{\rightmargin 0pt} 
         \setlength{\itemsep}{0pt} \settowidth          
        {\labelwidth}{#1.}\sloppy}}{\end{list}}
 
\newcounter{itemlistc}
\newcounter{romanlistc}
\newcounter{alphlistc}
\newcounter{arabiclistc}

\newcommand{\fcaption}[1]{
        \refstepcounter{figure}
        \setbox\@tempboxa = \hbox{\eightrm Fig.~\thefigure. #1}
        \ifdim \wd\@tempboxa > 5in
           {\begin{center}
        \parbox{5in}{\eightrm \smalllineskip Fig.~\thefigure. #1 }
            \end{center}}
        \else
             {\begin{center}
             {\eightrm Fig.~\thefigure. #1}
              \end{center}}
        \fi}
 
\newcommand{\tcaption}[1]{
        \refstepcounter{table}
        \setbox\@tempboxa = \hbox{\eightrm Table~\thetable. #1}
        \ifdim \wd\@tempboxa > 5in
           {\begin{center}
        \parbox{5in}{\eightrm\smalllineskip Table~\thetable. #1 }
            \end{center}}
        \else
             {\begin{center}
             {\eightrm Table~\thetable. #1}
              \end{center}}
        \fi}
 
\def\@citex[#1]#2{\if@filesw\immediate\write\@auxout    
        {\string\citation{#2}}\fi                       
\def\@citea{}\@cite{\@for\@citeb:=#2\do                 
        {\@citea\def\@citea{,}\@ifundefined             
        {b@\@citeb}{{\bf ?}\@warning
        {Citation `\@citeb' on page \thepage \space undefined}}
        {\csname b@\@citeb\endcsname}}}{#1}}
 
\newif\if@cghi
\def\cite{\@cghitrue\@ifnextchar [{\@tempswatrue
        \@citex}{\@tempswafalse\@citex[]}}
\def\citelow{\@cghifalse\@ifnextchar [{\@tempswatrue
        \@citex}{\@tempswafalse\@citex[]}}
\def\@cite#1#2{{$\null^{#1}$\if@tempswa\typeout
        {IJCGA warning: optional citation argument
        ignored: `#2'} \fi}}

\def\pmb#1{\setbox0=\hbox{#1}
        \kern-.025em\copy0\kern-\wd0
        \kern.05em\copy0\kern-\wd0
        \kern-.025em\raise.0433em\box0}

\def\fnt#1#2{\footnotetext{\kern-.3em
        {$^{\mbox{\scriptsize #1}}$}{#2}}}
 
\def\fpage#1{\begingroup
\voffset=.3in
\thispagestyle{empty}\begin{table}[b]\centerline{\footnotesize #1}
        \end{table}\endgroup}
 
\def\runninghead#1#2{\pagestyle{myheadings}
\markboth{{\eightit{\quad #1}}\hfill}{\hfill{\eightit{#2\quad}}}}
 
\font\tenrm=cmr10
\font\tenbf=cmbx10
\font\tenit=cmti10
\font\tenit=cmti10
\font\bfit=cmbxti10 at 10pt
 1
 1
 1

\font\ninerm=cmr9

\font\eightrm=cmr8
\font\eightit=cmti8

\def\qed{\hbox{${\vcenter{\vbox{                          
   \hrule height 0.4pt\hbox{\vrule width 0.4pt height 6pt
   \kern5pt\vrule width 0.4pt}\hrule height 0.4pt}}}$}}

\runninghead{J.S.S\'a Martins and P.M.C.de Oliveira}{Lattice Simulation of
Nuclear Multifragmentation}

\begin{document}
\normalsize\textlineskip
{\thispagestyle{empty}
\setcounter{page}{1}
 
\renewcommand{\thefootnote}{\fnsymbol{footnote}} 
\def\bsc{{\sc a\kern-6.4pt\sc a\kern-6.4pt\sc a}}
\def\bflatex{\bf L\kern-.30em\raise.3ex\hbox{\bsc}\kern-.14em
T\kern-.1667em\lower.7ex\hbox{E}\kern-.125em X}
 
\copyrightheading{Vol. 0, No. 0 (1997) 000--000}
 
\vspace*{0.88truein}

\def\OR{\hbox{\vrule height 6pt depth -3pt \vrule height 1pt depth 2pt}
\hskip 2pt}
 
\fpage{1}
\centerline{\bf LATTICE SIMULATION OF NUCLEAR MULTIFRAGMENTATION}
\vspace{0.37truein}
\centerline{\footnotesize J.S. S\'A MARTINS and P.M.C. DE OLIVEIRA}
\vspace*{0.015truein}
\centerline{\footnotesize\it  Instituto de F\'{\i}sica, Universidade
Federal Fluminense}
\baselineskip=10pt
\centerline{\footnotesize\it Av Litor\^anea, s/n, Boa viagem, Niter\'oi, RJ, 24210-340, Brazil}
\vglue 10pt
\centerline{\footnotesize\it  e-mail: jssm@if.uff.br and pmco@if.uff.br}
\vglue 10pt
\publisher{(received date)}{(revised date)}
 
\vspace*{0.21truein} 
\abstracts{\noindent Motivated by the decade-long
debate over the issue of criticality supposedly observed in nuclear
multifragmentation, we propose a dynamical lattice model to simulate the
phenomenon. Its Ising Hamiltonian mimics a short range attractive 
interaction which competes with a thermal-like dissipative process. The
results here presented, generated through an event-by-event analysis, are
in agreement with both experiment and those produced by a percolative
(non dynamical) model.  }{}{}
 
\vspace*{1pt} 
\textlineskip 
\section{Introduction} 
\vspace*{-0.5pt}
\noindent 
\textheight=7.8truein 
\setcounter{footnote}{0}
\renewcommand{\thefootnote}{\alph{footnote}} 

In the early 1980's, the experimental observation of power law-like
distribution of fragment yields in nuclear reactions induced by high
energy proton projectiles~\cite{Wadd,Finn} initiated an ongoing
controversy on the origins of this behavior. When a proton or a heavy ion
collides with a heavy nucleus, the target can break into a number of
nuclei with atomic number $Z>2$ in addition to several nucleons and alpha
particles. The number of heavy fragments induces a classification of these
processes into three different regimes: {\it fission}, with two heavy
fragments; {\it spallation}, yielding one heavy fragment and a few light
ones; and {\it multifragmentation}, where the product includes no heavy
fragments, but is composed of several light nuclei of varying
sizes~\cite{Hufner}. There is general agreement that the boundary between
light and heavy fragments lies around $Z=20$ \cite{Panagiotou}.  The
processes of fission and spallation are quite well understood, but such is
not the same with multifragmentation. One of the reasons for this lack of
knowledge stems from experimental difficulties - a detailed investigation
of fragmentation requires coincident measurements of the multiple
fragments formed. Most of the experiments reported in the literature
measure little more than the inclusive mass yield of fragments - and so
cannot make a clear distinction between those that correspond to each kind
of phenomenon involved. Most of the exclusive experiments, on the other
hand, are emulsion experiments with inverse kinematics~\cite{Wadd}, and
suffer the drawback of poor statistics. The lack of reliable coincidence
data has forced theoretical investigations to concentrate on the
explanation of the mass yield curve. 

For $Z\leq 20$, it has been pointed out that this curve is compatible with
a power-law dependence for the cross-section $\sigma \sim Z ^{-\tau} $,
where $\tau \sim 2.2$ is practically independent of the beam energy and
the exact composition of projectile and target. This parametrization of
the mass yield curve, along with the measured value of the exponent, is
precisely what is expected for the transition between liquid and gaseous
phases in nuclear matter \cite{Panagiotou} - this transition takes its
name from a parallel with the process of droplet formation in a similar
transition for fluids \cite{Fisher}. This fact has been pointed out as a
clear indication that multifragmentation is strongly related to that phase
transition. 

The currently accepted scenario for the fragmentation of
nuclei begins with a quick isentropic expansion, as the target nucleus is
hit \cite{Ngo}. The resulting compound nucleus has its density diminished
in this process, and the fragmentation begins with a cracking of the
system. In this way, primary fragments are formed, still highly excited.
Deexcitation of these fragments is acompanied by the formation of
secondary fragments \cite{Donan}. During the whole process, the system has
enough time for thermal equilibration; we have two different time-scales
involved - namely, in this case, we have fast formation of cracks and slow
thermalization -, one of the known conditions for the installation of a
complex regime. The theoretical understanding of this process is very
difficult, since we are in the presence of a true many-body problem, due
to long range correlations between the clusters that are being formed.

Among the theoretical models that have been proposed to study this
phenomenon ( for a brief review, see Ref.[\cite{Caio}] ), by far the most
popular one is a bond-percolation model on a cubic lattice \cite{Bauer}.
It has been shown that, independently of the knowledge of the parameter
that would control the approach to criticality, a number of successful
comparisons could be made between predictions of this model and
experimental results \cite{Campi86}. By studying the moments of the
cluster size distributions on an event-by-event analysis, and assuming a
scaling property for these distributions \cite{Stauffer}, one can show
that a strong correlation between these moments must be present in the
vicinity of the critical region. Nevertheless, the purely geometrical and
static character of such a model brings about the need for one that could
rely on some dynamical interaction between the nucleons participating in
the process.\\ 

In the next section, we present a dynamical dissipative
lattice model that responds to that need. Some details of its computer
implementation are then discussed. The last section is dedicated to a
presentation of our results, and to a comparison between them and those
coming from the percolative model. 

\section{The Model} 
\noindent 

Initial versions of the model we present in
this paper were used in the context of surface wetting \cite{Manna} and of
drop formation on the leaky faucet problem \cite{Murilo,Tadeu}. Its use in
nuclear collisions was foreseen in the analysis of residual mass in the
evaporation of hot nuclei \cite{Eu}. The success obtained in that
application encouraged us to the extension presented here. 

The nucleus,
in its initial configuration, is a dense cluster of occupied cells in an
otherwise empty cubic lattice - using the terminology of magnetic systems
in the context of the gas-lattice model, a cluster of spins up surrounded
by spins down - subject to an Ising-type interaction given by the model
Hamiltonian \begin{equation} {\cal H} = - J\sum_{<i,j>} S_{i}S_{j},
\label{H} \end{equation} where the summation extends to nearest and
next-nearest neighbours and $S_{i,j}= \pm 1$. With this last prescription,
we intend to mimic a surface tension, having in mind a liquid drop model
for the nucleus. The initial excitation energy of this compound nucleus -
target plus absorbed projectile - is associated to a temperature
parameter, which will control the transition probability between two
different configurations of the system. The nucleus is then subjected to a
dissipative Metropolis dynamics \cite{Metropolis}, generating a Markovian
chain of configurations. This dynamics involves:\\ 
(i) a random choice of an occupied site at the cluster perimeter, followed
by a random choice of an unoccupied site, not contiguous to the first,
also at the perimeter;\\
(ii) a double flip of the spins at these two sites, decided by the
Metropolis rule - thus involving the temperature parameter already
mentioned;\\ 
(iii) if the double flip occurs, it is verified if this
configuration is still a connected set of spins up; if it is not, we have
the formation of a primary fragment, whose mass is contabilized, and energy
 is dissipated, in the form of a decrease in temperature
(intermittent, fast cracking energy dissipation);\\ 
(iv) in either case, an additional small decrement
in temperature is promoted, simulating the emission of radiation
by the system (continuous, slow thermalization). 

The recently formed fragment is erased, and this dynamics continues until
a low temperature is attained, when no more fragments would be formed.\\ 

The distribution of fragments
thus obtained is characterized by its moments. If $n_{i}(s)$ is the number
of fragments with mass $s$ obtained in the event $i$, its $k^{th}$ moment
is computed by $$M_{i}(k) = \sum_{s} s^{k} n_{i}(s).$$\\
In a normal
thermodynamic system, these moments would diverge in the thermodynamic
limit at the critical point, for $k>1$, with critical exponents
$$M(k,\epsilon) \sim \epsilon ^{-\mu_{k}},$$ where $\epsilon$ measures the
distance to the critical point. If the distribution $n(s)$ has the scaling
property \cite{Fisher,Stauffer} $$n(s,\epsilon)\sim
s^{-\tau}f(\epsilon s^{\sigma}),$$where $\tau$ and $\sigma$ are two
critical exponents, one can show that moments of different orders must be
correlated. The relation between the exponents of moments divergence and
those of the distribution is given by \cite{Campi86}
$$\mu_{k}=-(\tau-1-k)/\sigma .$$\\ 
Since we are working with intrinsically
finite systems - the nuclei - the moments will remain finite, even for
$k>1$. The normal signature of critical behavior - the divergence of the
moments of the distribution - is washed out by the finiteness of the
system. This is one of the difficulties that faces theoretical work in
small systems. Nevertheless one can use the correlation between moments of
different orders to examine the surviving traces of critical behavior.

It is more natural to work with normalised moments
$$S_{i}(k)=M_{i}(k)/M_{i}(1)$$ instead of the regular ones. Then it can be
shown \cite{Campi86} that 
$$\hbox{log}(S_{i}(3))=\lambda_{3/2}\hbox{log}(S_{i}(2))$$ and
$$\hbox{log}(S_{i}(5))=\lambda_{5/2}\hbox{log}(S_{i}(2) )$$ 
where, with the usual identification $\gamma=\mu_{2}$,
$$\lambda_{3/2}=1+1/\sigma \gamma ,$$ and $$\lambda_{5/2}=1+3/\sigma
\gamma .$$

\section{Computer Strategy} 
\noindent 

We will adopt the C programming
language sintax~\cite{KR} in this section, with some minor explanations. 
Our strategy follows two rules: storing data (spin states) bit by bit on
computer words; and treating them almost exclusively by bitwise logical
operators. Obviously, this saves computer memory by a factor of 32 (for 32
bit word computers). Better yet, computer time can be saved by a similar
factor whenever parallel updating is possible; such is the case in some of
the routines discussed below. Many general tricks designed to implement
this strategy, similar to the particular ones used here, are discussed in
detail in references~[\cite{Tadeu,PM}]. 

Our model nucleus resides on a 32x32x32 cubic lattice. This lattice is
mapped onto a vector L[$r$] ($r = 0, 1, 2 \dots lastword$), that keeps the
current nucleus shape, through the rule: element ($i,j,k$) is represented
by the $k^{th}$ bit of L[$j*32 + i$] ($lastword$ is a mnemonic for
$32*32-1$). The rationale behind this choice lies in the fact that,
although a mathematical operation is needed whenever a particular cell
must be accessed, that is exactly what is done at machine code level - so
very little extra time is spent. On the other hand, most of the really
time-consuming operations are done on a sequential basis; in these cases,
storing the array as a single-index vector avoids the need to go through
multiple accesses by pointers, which is the way arrays are usually
implemented on computers. 

As an example of all that has just been said, we write down the algoritm
used to determine the inner nucleus boundary. This boundary is stored on
array I[$r$]
by
\vskip .5cm
for($r=0; r<=lastword; r++$) \{\par

\hskip .5cm
 I[$r$] = ($\sp\sim$L[$r$]$<<$1) \OR ($\sp\sim$L[$r$]$>>$1) \OR
 $\sp\sim$L[$r-1$] \OR $\sp\sim$L[$r+1$] \OR $\sp\sim$L[$r+32$] \OR $\sp\sim$L[$r-32$]; /* first neighbors */\par
\hskip .5cm
 I[$r$] = I[$r$] \OR $\sp\sim$L[$r-1-32$] \OR $\sp\sim$L[$r-1+32$] \OR
$\sp\sim$L[$r+1-32$] \OR $\sp\sim$L[$r+1+32$] \OR ($\sp\sim$L[$r-1$]$<<$1)
\OR ($\sp\sim$L[$r-1$]$>>$1) \OR ($\sp\sim$L[$r+1$]$<<$1 \OR
($\sp\sim$L[$r+1$]$>>$1) \OR ($\sp\sim$L[$r-32$]
$<<$1) \OR ($\sp\sim$L[$r-32$]$>>$1) \OR ($\sp\sim$L[$r+32$]$<<$1) \OR
($\sp\sim$L[$r+32$]$>>$1); /* and second neighbors */
\par \hskip .5cm
I[$r$] = I[$r$] \& L[$r$]; \}
\par \vskip .5cm
\noindent where $\sp\sim$,
\OR and \& represent respectively the bitwise logical operations NOT, OR
and AND. A similar procedure gives the outer drop boundary.  A pair of
non-adjacent positions are randomly selected, one from each of these
boundaries, as candidates for a mass-conserving, Kawasaki-like updating.
The energy variation involved in this exchange of positions is also
calculated through the use of bitwise logical operations. 

The determination of the connectedness of the nucleus after the spin
updating is a dificult problem, due to its non-locality. There is no clear
shortcut solution in the literature for three dimensions, and here resides
the most time-consuming algorithm in the program. We chose the so-called
burning algorithm~\cite{burning}, designed to determine connected clusters
of lattice sites sharing some property. A computer implementation for a
similar problem in two dimensions was published in Ref.[\cite{Tadeu}]. 

\section{The Results}
\noindent

For a close examination of the correlation between normalised moments of
different orders, we made log-log plots of $S(3) \times S(2)$ and $S(5)
\times
S(2)$, following the suggestions originally made by Campi \cite{Campi86}. 
In fig. 1 and 2 we show the evolution of these correlations with the
initial excitation energy of the system. It is clear from these graphs
that the model reproduces the strong correlation observed experimentally,
as also did the percolation model. The points making an apparent arc of
circle near the origin for low excitation energies in both instances are
themselves also present in the experiments, and are probably related to
very 'gentle' events, yielding mostly fragments of small mass ($<4$). The
measured slopes $\lambda_{3/2}=2.28\pm 0.01$ and $\lambda_{5/2}=4.75\pm
0.04$ agree with both experiment and percolation model \cite{Campi86}.

\begin{figure}[!h]
\vspace {0.2in}
\epsfysize=2.0truein
\centerline{\epsfbox{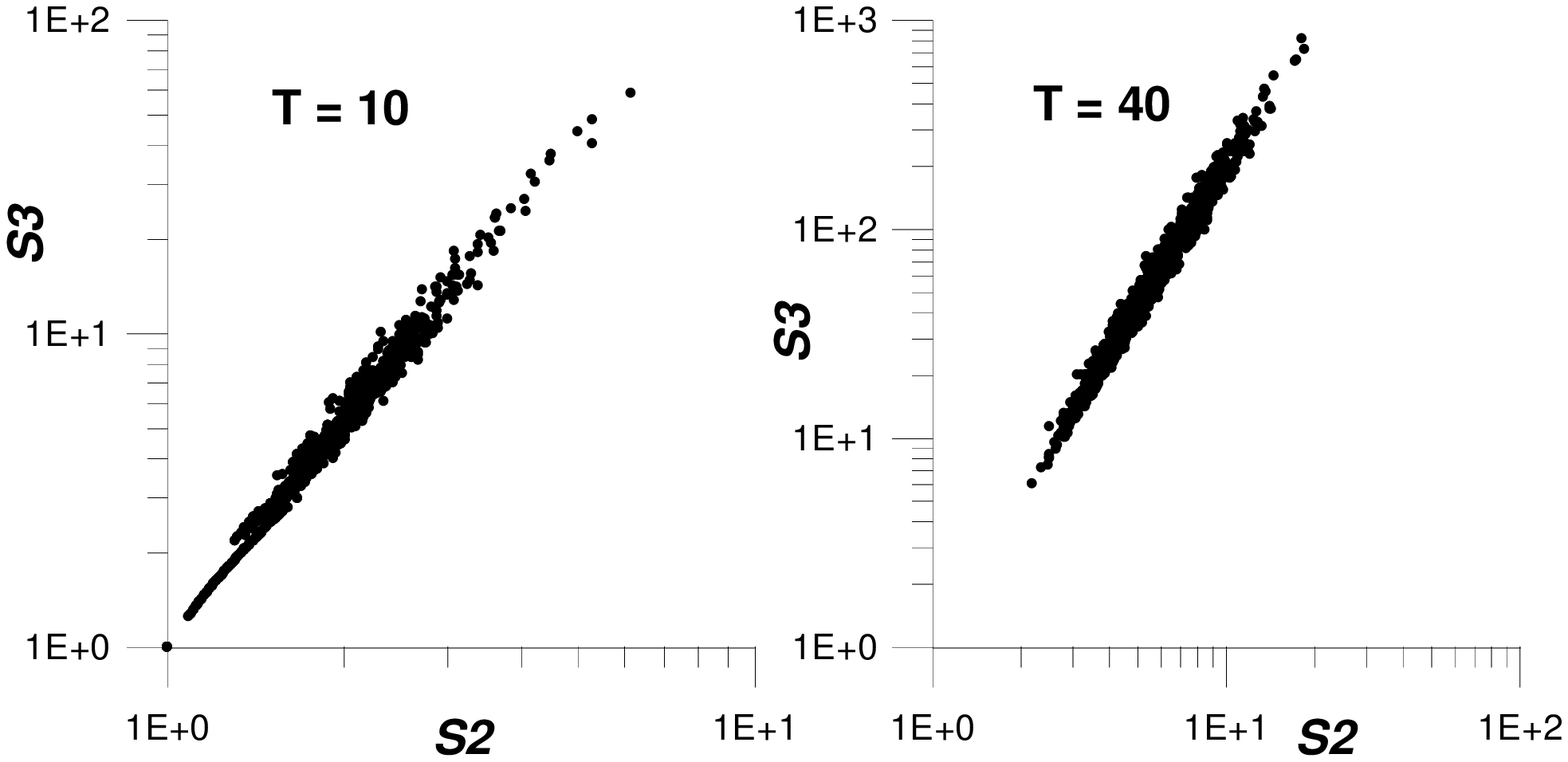}}
\fcaption{Plot of $\hbox{log}(S_{3}) \times \hbox{log}(S_{2})$. The system
used in simulations is a 6x6x6 cubic cluster - total number of "nucleons"
is 216, permitting close comparisons with the results quoted by
Ref.[\cite{Campi86}].}
\end{figure}

\begin{figure}[!h]
\vspace {0.2in}
\epsfysize=2.0truein
\centerline{\epsfbox{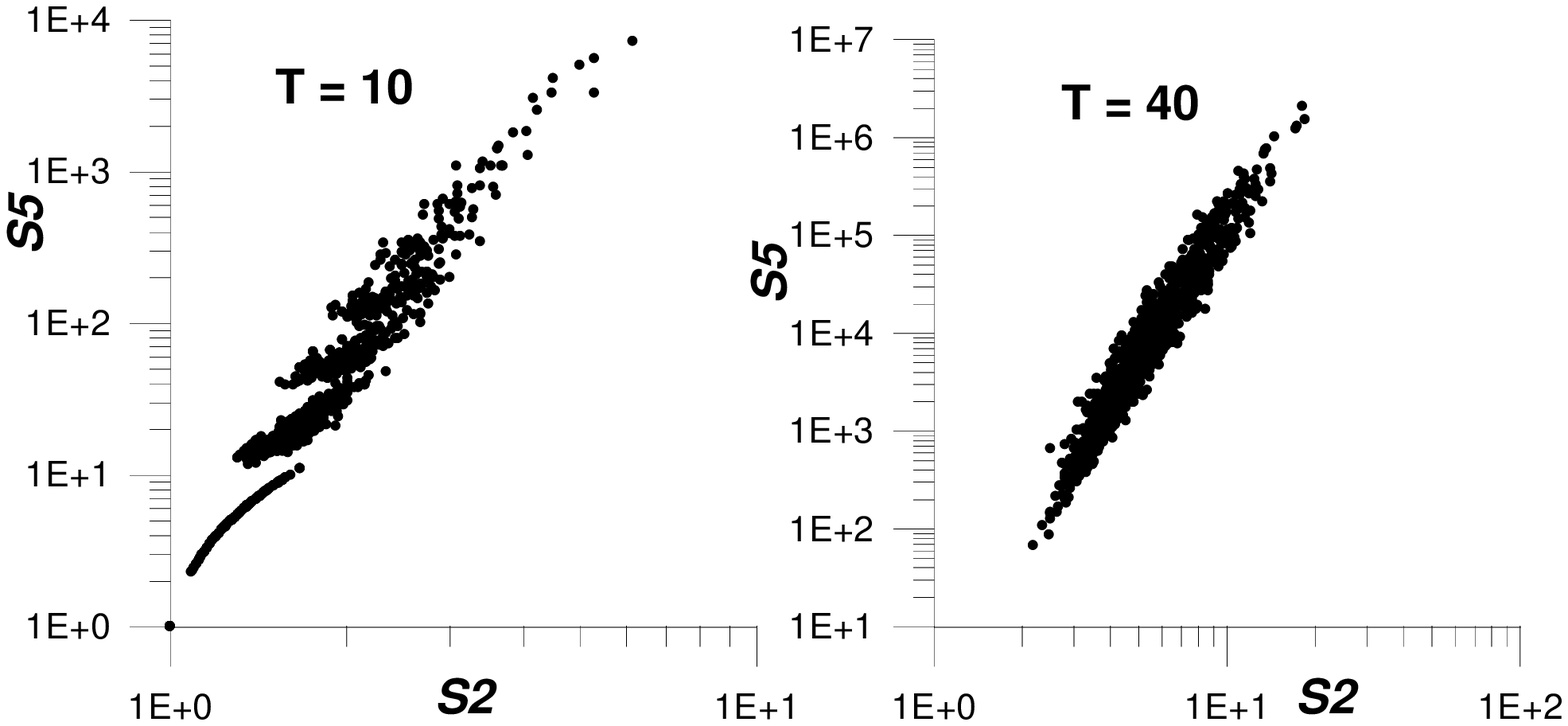}}
\fcaption{Plot of $\hbox{log}(S_{5}) \times \hbox{log}(S_{2})$. The points
closer to the origin on the graph on the left correspond to highly
undercritical events, as explained in the text.}
\end{figure}

Observation of the data gives an indication that our model is probably
more 'selective' in the energy involved in the reaction then the
experiments they are being compared with - the data used was from
inclusive collisions \cite{Wadd}. A selection of the initial temperature
parameter can bring the system closer to the critical region. This makes
it a candidate for modelling the new exclusive experiments now being
made. 

The determination of the proximity to the critical region made use
of the so called 'Campi scatter plot' (fig.3). This is a log-log plot of
the greatest fragment in each event against the second normalized moment,
averaged over events with the same $M(1)$. In this plot one can identify
three different regions \cite{Mastinu}: a region with negative slope,
corresponding to events with small energy - or {\it undercritical} -, a
region with positive slope, or {\it overcritical} and a region where there
is a great dispersion in the second moment. We understand that this is the
critical regime we are looking for. In fig.3 we show the evolution of this
plot with the increase in the temperature parameter. The simulation with
$T=40$ was chosen as a representative of the experimental situation, and
determined the temperature where the slopes $\lambda_{3/2}$ and
$\lambda_{5/2}$ were measured. The measured slope of the scatter plot in
the undercritical region, although very inaccurate, is also coherent with
the result of the percolative model.

\begin{figure}[!h] 
\epsfysize=4.0truein
\centerline{\epsfbox{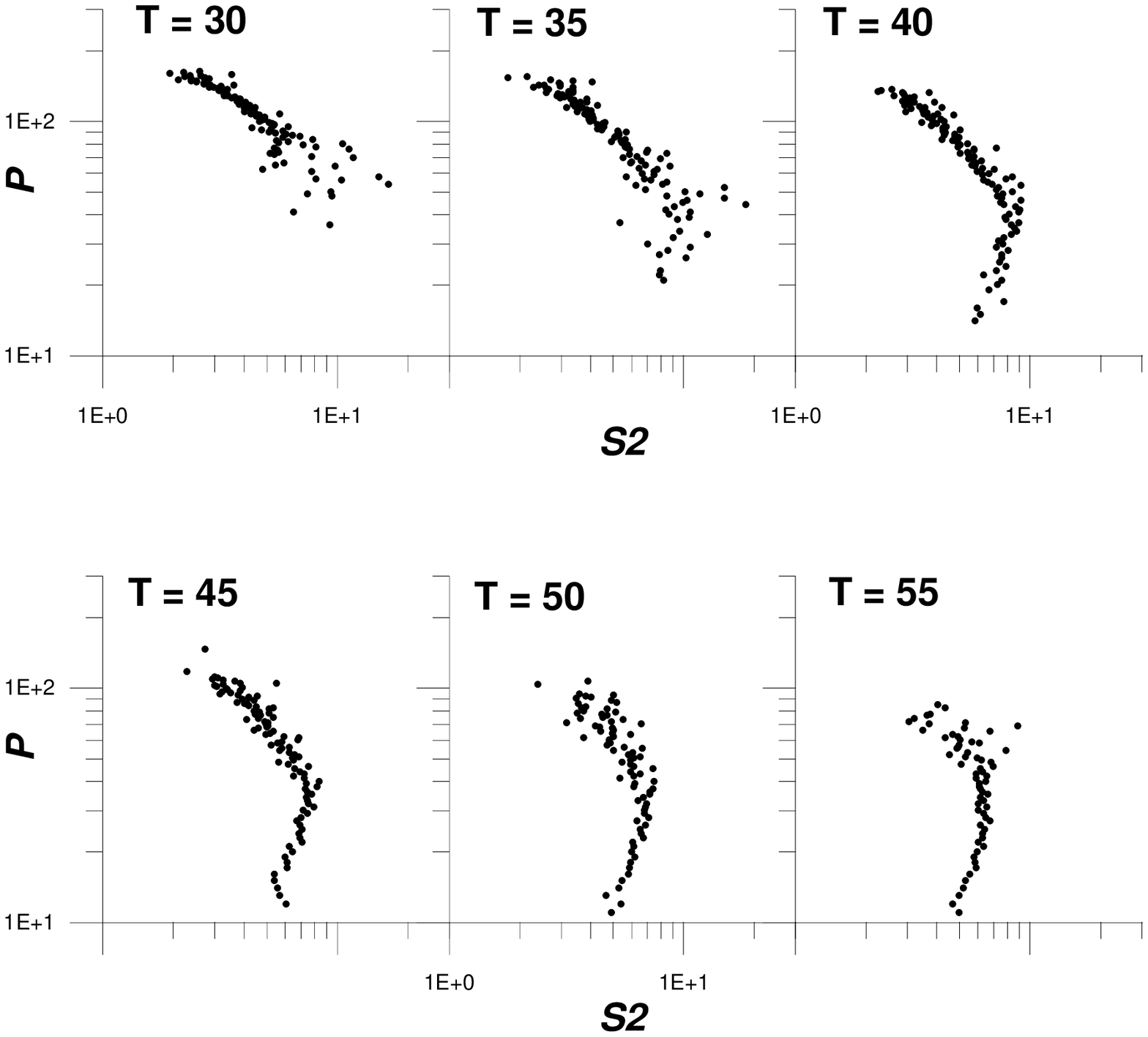}}
\fcaption{Evolution of Campi's Scatter Plot with the initial value of the
temperature parameter. As this value is increased, the system goes more
and more into the critical regime, loosely associated with events
represented by points at the intersection of the negative and positive
slopes portions of the graph.} 
\end{figure}

In summary, we showed that a dissipative dynamic model can reproduce much
of the observed behaviour of the distribution of fragments produced in
nuclear reactions induced by collisions with high energy protons or heavy
ions. These results allow us to conclude that a few model dynamical
ingredients, namely a short range nucleon-nucleon attraction, fast energy
dissipation during each crack formation and continuous, slow
thermalization are enough: a detailed description of the microscopic
interactions is not needed. A complete characterisation of the fragment
distribution generated through the use of this model is now in progress.\\

%
 
\nonumsection{Acknowledgements}
\noindent

We would like to thank S.M. de Oliveira, T.J.P.Penna, and A. Delfino for
helpfull discussions. This work was partially suported by Brazilian
agencies CNPq, FINEP, CAPES and FAPERJ.

\vfill

 
\end{document}